\documentclass[fleqn,usenatbib]{mnras}

\usepackage{newtxtext,newtxmath}

\usepackage[T1]{fontenc}

\DeclareRobustCommand{\VAN}[3]{#2}
\let\VANthebibliography\thebibliography
\def\thebibliography{\DeclareRobustCommand{\VAN}[3]{##3}\VANthebibliography}

\usepackage{graphicx}	
\usepackage{amsmath}	

\title[Shock Driven Mass Loss]{Partial Stellar Explosions - Ejected Mass and Minimal Energy}

\author[Linial, Fuller \& Sari]{
Itai Linial,$^{1}$\thanks{E-mail: itai.linial@mail.huji.ac.il}
Jim Fuller,$^{2}$
Re'em Sari$^{1}$
\\
$^{1}$Racah Institute of Physics, The Hebrew University, Jerusalem 91904, Israel\\
$^{2}$TAPIR, Mailcode 350-17, California Institute of Technology, Pasadena, CA 91125, USA
}

\date{Accepted XXX. Received YYY; in original form ZZZ}

\pubyear{2020}

\begin{document}
\label{firstpage}
\pagerange{\pageref{firstpage}--\pageref{lastpage}}
\maketitle

\begin{abstract}
 Many massive stars appear to undergo enhanced mass loss during late stages of their evolution. In some cases, the ejected mass likely originates from non-terminal explosive outbursts, rather than continuous winds. Here we study the dependence of the ejecta mass, $m_{\rm ej}$, on the energy budget $E$ of an explosion deep within the star, using both analytical arguments and numerical hydrodynamics simulations. Focusing on polytropic stellar models, we find that for explosion energies smaller than the stellar binding energy, the ejected mass scales as $m_{\rm ej} \propto E^{\varepsilon_{m}}$, where $\varepsilon_m = 2.4-3.0$ depending on the polytropic index. The loss of energy due to shock breakout emission near the stellar edge leads to the existence of a minimal mass-shedding explosion energy, corresponding to a minimal ejecta mass. For a wide range of progenitors, from Wolf-Rayet stars to red supergiants, we find a similar limiting energy of $E_{\rm min} \approx 10^{46}-10^{47} \rm \, erg$, almost independent of the stellar radius. The corresponding minimal ejecta mass varies considerably across different progenitors, ranging from $\sim \! 10^{-8} \, \rm M_\odot$ in compact stars, up to $\sim \! 10^{-2} \, \rm M_\odot$ in red supergiants. We discuss implications of our results for pre-supernova outbursts driven by wave heating, and complications caused by the non-constant opacity and adiabatic index of realistic stars.
\end{abstract}




\section{Introduction}
Massive stars appear to shed large amounts of mass during late stages of their evolution. In many cases, episodic eruptions, rather than continuous line-driven winds, are the dominant source of mass loss from these stars \citep{Smith_2014}. This property of massive stars is a fundamental ingredient in a variety of astronomical phenomena - type IIn and Ibn supernovae (SNe) are powered by the collision of core-collapse supernova ejecta with dense circumstellar environments formed by earlier eruptive mass loss; giant eruptions of luminous blue variables (LBVs) involve the ejection of large amounts of stellar material (with the best-studied example being $\eta$ Carinae, surrounded by 10-20 $\rm M_\odot$ of ejecta e.g., \citealt{Smith_2013}). 

While observational evidence of eruptive mass loss in massive stars is steadily growing, the theoretical understanding of these eruptions is still incomplete. Several mechanisms have been proposed as the trigger of these stellar eruptions - unsteady nuclear burning \citep{Smith_Arnett_2014}, pulsational pair instability \citep{Woosley_2002,Woosley_2007}, stellar collisions in a binary system \citep{Podsiadlowski_2010}, or wave-driven mass loss \citep{quataert:12,shiode:14,fuller:17,fuller:18}. All of these mechanisms involve energy deposition deep within the stellar envelope, resulting in a mass-shedding eruption. Depending on the duration of energy injection relative to the dynamical timescale, mass loss can be either continuous, driven by super-Eddington winds \citep{quataert:16}, or instantaneous, as a result of an explosion driving a shock wave through the stellar envelope \citep{Smith_2014}. 

In this paper, we focus on eruptive mass loss driven by non-terminal explosions. We study the evolution of a thermal explosion set deep within the stellar interior, its deceleration as a Sedov-Taylor explosion, its spherical expansion as a sound pulse, its steepening to a shock front (see \citealt{Ro_Matzner_2017} for detailed study of the steepening phase), its transition to a strong shock during its approach towards the stellar surface, and finally the resulting mass ejection and the termination of shock acceleration as it breaks out from the surface. Without adhering to any specific explosion mechanism, we try to draw general conclusions relevant for a wide range of progenitor stars. By studying polytropic stellar models, our analytical and numerical results produce simple power-law scaling relations, that can be easily extended to arbitrary progenitor mass and radius. 

Our analytical treatment is reminiscent of theoretical works that study the dynamics of supernova shocks and their breakout \citep[e.g.,][]{Matzner_McKee_1999,Nakar_Sari_2010}, with the key difference being the explosion energy scale. While the typical supernova energy scale is $\sim 10^{51} \rm erg$, here we focus on explosion energies smaller than the star's binding energy, thus resulting in partial mass ejection, and involving stages where the shock wave propagates as a weak shock.

A few authors have considered similar energy scales and studied different aspects of eruptive mass loss events. \cite{Dessart_2010} numerically investigated the response of stellar envelopes to sudden energy deposition and obtained the resulting luminosity. \cite{Owocki_2019} studied energy injection at different locations within the stellar interior, and obtained the density and velocity distribution of the resulting mass ejection. \cite{Kuriyama_Shigeyama_2020} carried out 1D-radiation hydrodynamical simulations to obtain the amount of ejecta mass and compute the light curve associated with pre-SN outbursts.

A primary goal of this work is to find the \textit{minimal} energy budget required for mass ejection, and find how the amount of ejected mass scales with the explosion energy. The existence of a minimal explosion energy (and correspondingly, minimal ejecta mass) arises from the finite width of radiative shocks accelerating in the decreasing density profile near the stellar edge. Shock acceleration terminates at the breakout layer - where the shock's width is comparable to its distance from the edge. Upon breakout, if the shock speed is not comparable to or greater than the escape speed, no stellar material becomes unbound. 

The paper is organized as follows. In section \ref{sec:min_energy_general} we discuss the theoretical lower limit on shock energy that results in mass ejection from a star, and provide analytical estimates for polytropes. In section \ref{sec:Shocks_from_core} we consider the mass ejection resulting from a point explosion in the stellar core, deriving analytical results that depend on the explosion energy. The results of numerical hydrodynamics simulations and their application to typical progenitor stars are presented in section \ref{sec:numerical_investigations}. We discuss our results and summarize our conclusions in section \ref{sec:discussion}.

\section{Minimal energy required for mass ejection} \label{sec:min_energy_general}
Consider a star of mass $M_\star$ and radius $R_\star$. When energy $E_{\rm dep}$, much larger than $E_{\rm bind} \approx GM_\star^2 / R_\star$ is deposited within it, the star {is completely destroyed, with no self-gravitating remnant surviving the explosion}. If however $E_{\rm dep}<E_{\rm bind}$, just a fraction of the stellar envelope becomes unbound. Here we estimate the minimal amount of energy that results in partial mass ejection.

The injected energy produces an outward propagating spherical shock wave, accelerating as it approaches the steep density gradient near the stellar edge. Material accelerated by the shock to beyond the star's escape velocity becomes unbound.

If the shock is radiation mediated, its acceleration terminates when its distance from the stellar edge becomes comparable to the shock's finite width. This layer (the \textit{breakout} layer) is characterized by the condition \begin{equation} \label{eq:tau_bo}
    \tau_{\rm bo} = \frac{c}{3v_{\rm sh,bo}} \,,
\end{equation}
where $\tau_{\rm bo}$ is the optical depth measured from the stellar surface inwards, and $v_{\rm sh,bo}$ is the shock velocity at this layer. The material velocity long after the shock has passed is similar to the shock velocity, $\delta v_{\rm f} = C_{\rm f} v_{\rm sh}$, where $C_{\rm f} \approx 2$ (e.g., \citealt{Sakurai_1960}, \citealt{Matzner_McKee_1999}, \citealt{Ro_Matzner_2013}). Thus, the shock velocity at the breakout layer must satisfy $v_{\rm sh,bo} > v_{\rm esc}/C_{\rm f}$ in order to allow for mass ejection to occur, and in the limiting case, the breakout layer's optical depth is
\begin{equation} \label{eq:tau_bo_limit}
    \tau_{\rm bo} = \frac{C_{\rm f}}{3} \frac{c}{v_{\rm esc}} \,.
\end{equation}
Note that if $v_{\rm sh,bo} < v_{\rm esc}/C_{\rm f}$, the shock breaks out and dissipates without accelerating \textit{any} material to beyond the star's escape velocity. This criterion therefore distinguishes between shocks that result in mass ejections, and shocks that do not expel any material from the star. If the opacity $\kappa$ is constant in the star's outer layers, the optical depth $\tau_{\rm bo}$ sets the minimal ejecta mass
\begin{equation} \label{eq:m_bo}
    m_{\rm bo} = 4\pi \frac{R_\star^2}{\kappa} \tau_{\rm bo} = \frac{4\pi C_{\rm f}}{3} \frac{R_\star^2}{\kappa} \frac{c}{v_{\rm esc}} = \frac{4\pi C_{\rm f}}{3\sqrt{2}} \left( \frac{c^2 R_\star^5}{G \kappa^2 M_\star} \right)^{1/2} \,,
\end{equation}
corresponding to the limiting case.

As the shock accelerates within the decreasing density of the outer stellar layers, it evolves following a second-kind self-similarity solution found by \cite{GFK_1956} and \cite{Sakurai_1960} (hereby GFKS). These self-similarity solutions pass through a sonic-point, and the shock front is therefore causally disconnected with the shocked downstream, implying that the shock's energy content decreases as it approaches the surface. Most of the shock energy remains at the layer at which the shock started to accelerate, whereas the energy at the breakout layer is just a minute fraction of the total shock energy. We continue by identifying the layer at which the shock has started to accelerate as a GFKS shock, in order to estimate the minimal energy deposition, $E_{\rm dep,min}$.

Following \cite{Sakurai_1960}, extrapolating inwards from the breakout layer, the shock velocity scales as
\begin{equation} \label{eq:v_sh_rho}
    v_{\rm sh}(\rho) = \frac{v_{\rm esc}}{C_{\rm f}} \left( \frac{\rho_{\rm bo}}{\rho} \right)^{\mu} \,,
\end{equation}
where $\rho$ is the (unperturbed) density evaluated just in front of the shock, $\rho_{\rm bo}$ is the density at the breakout layer, $\tau(\rho_{\rm bo})=\tau_{\rm bo}$ and $\mu \approx 0.19$ (see \citealt{Ro_Matzner_2013} for a thorough investigation of the value of $\mu$). 

\begin{figure}
    \centering
    \includegraphics[width=\columnwidth]{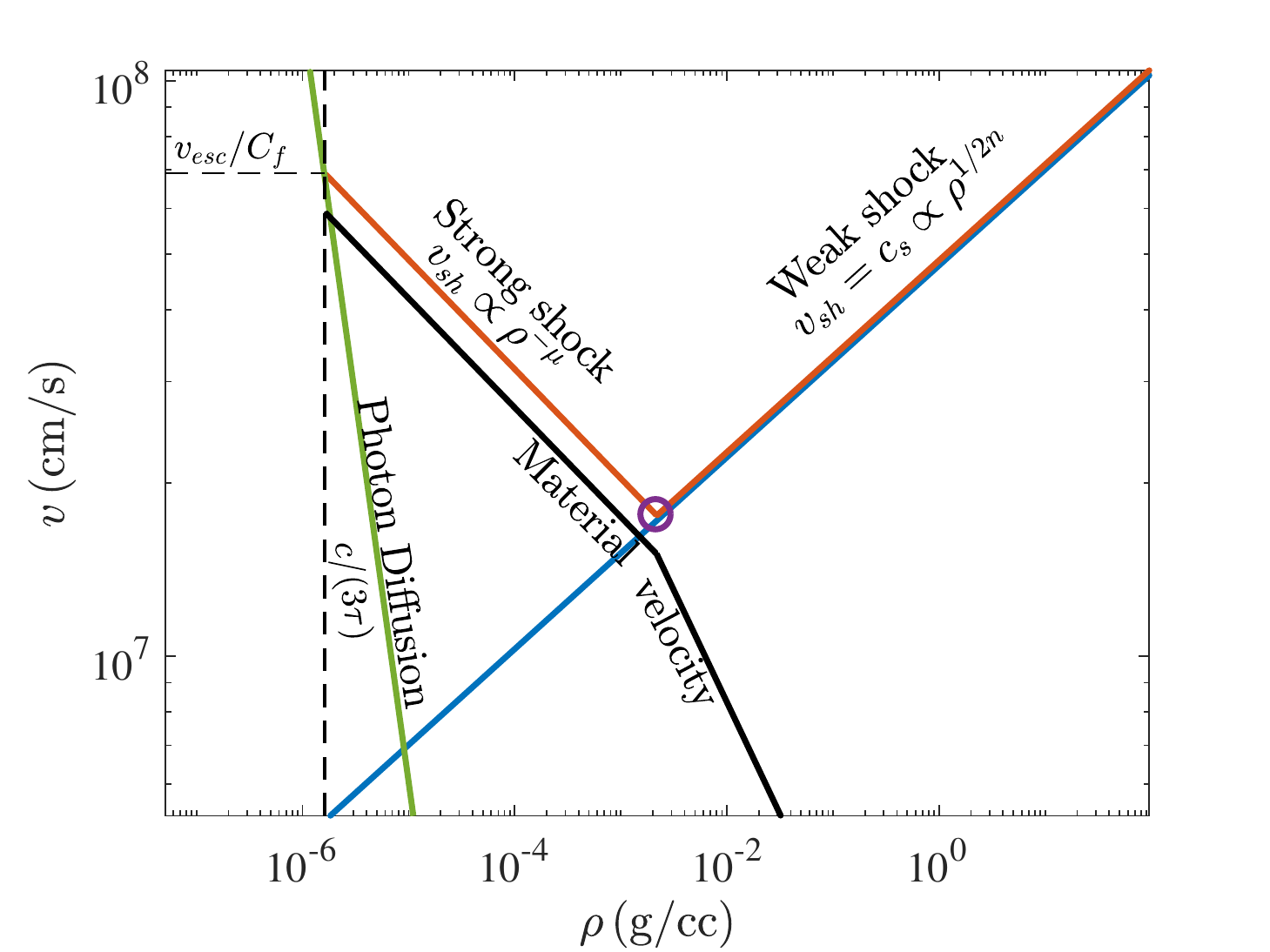}
    \caption{Shock velocity (\textit{red} curve) as a function of density in the limiting case, where mass ejection occurs, for a star of $M_\star = 10 \, \rm M_\odot$, $R_\star = 2 \, \rm R_\odot$, $n=3$ and $\kappa = 0.3 \, \rm cm^2 \, g^{-1}$. A weak shock propagates towards the stellar surface, traveling roughly at the local sound speed, plotted in \textit{blue}. As the density decreases, the post-shock material velocity (\textit{black} curve) increases until it becomes sonic at density $\rho_s$, marking the transition to a strong shock. At lower densities, the shock follows Sakurai's solution, described by equation \ref{eq:v_sh_rho}. The shock breaks out when its velocity becomes similar to the effective photon diffusion speed, $c/(3\tau)$, plotted in \textit{green}. In the limiting case, shown here, the shock velocity at breakout is $v_{\rm esc}/C_{\rm f}$, meaning that only the breakout layer becomes unbound. The shock's transition from weak to strong, highlighted by the \textit{purple circle} is also where most of the shock's energy is deposited (equation \ref{eq:E_dep_min}).}
    \label{fig:VisualizeMinShockEnergy}
\end{figure}

Sakurai's solution assumes a strong shock, i.e., negligible upstream pressure with respect to the downstream pressure, or equivalently, a shock that is very supersonic. Since the sound speed, $c_s$ generally decreases as $\rho$ decreases towards the surface of the star while $v_{\rm sh}$ increases, equation \ref{eq:v_sh_rho} becomes increasingly more accurate as the shock approaches the stellar edge. The density $\rho_s$ at which $v_{\rm sh}(\rho_s) = c_s(\rho_s)$ roughly marks the shock's transition from weak to strong. At deeper layers, where $\rho>\rho_s$, the shock is weak, and propagates at roughly the local speed of sound, $c_s(\rho)$ (see figure \ref{fig:VisualizeMinShockEnergy}).

The shock's transition from weak to strong, where it begins to accelerate, also defines the region in which most of the shock's energy is deposited. Hence, the minimal energy content in a weak shock that produces a mass-ejecting strong shock is, to within an order of magnitude
\begin{equation} \label{eq:E_dep_min}
    E_{\rm dep,min} \approx \delta m(\rho_s) c_s^2(\rho_s) \,,
\end{equation}
where $\delta m(\rho)$ is the mass enclosed in a scale height around density $\rho$. The energy $E_{\rm dep,min}$ is a lower limit on the amount of energy required to launch a shock that would unbind some of the stellar mass before breaking out at optical depth $\tau_{\rm bo}$. We summarize these arguments in figure \ref{fig:VisualizeMinShockEnergy}, where the shock's velocity is plotted as a function of density, for the limiting case. As we later show, when energy is instantaneously deposited in the stellar core, some of it is dissipated deep in the core, adding to the total required energy budget.

\subsection{Analytical Expressions for Polytropes} \label{sec:polytropes_analytical}

In this section, we derive analytical expressions assuming that the star is an $n$-polytrope. In the limit that both the breakout and the energy deposition layers are small compared with the stellar radius, simple power-law relations are obtained. In the following expressions, dimensionless, order unity prefactors are denoted by $K_X(n)$, as summarized in table \ref{tab:Polytropes_Constants}.

Hydrostatic equilibrium in polytropes implies that density scales as a power-law of the depth $x$, measured from the stellar edge inwards, assuming $x \ll R_\star$ as given by
\begin{equation} \label{eq:rho_x}
    \rho(x) = K_\rho (n) \frac{M_\star}{R_\star^3} \left( \frac{x}{R_\star} \right)^n \,,
\end{equation}
{where $n$ is related to the adiabatic index $\gamma$ by $\gamma = 1+1/n$, and $K_\rho(n)$ is a dimensionless coefficient given in table \ref{tab:Polytropes_Constants}.} The optical depth measured from the stellar surface is given by
\begin{equation}
    \tau = \kappa \int_{R_\star-x}^{R_\star}{\rho(x') \, dx'} = K_\tau(n) \frac{\kappa M_\star}{R_\star^2} \left( \frac{\rho}{M_\star/R_\star^3} \right)^{1+1/n} \,,
\end{equation}
where constant opacity $\kappa$ is assumed throughout the polytrope. The breakout density in the limiting case (equation \ref{eq:tau_bo_limit}) is thus
\begin{equation} \label{eq:rho_bo}
    \rho_{\rm bo} = K_{\rm \rho,bo}(n) \frac{M_\star}{R_\star^3} \left( \frac{G \kappa^2 M_\star^3}{c^2 R_\star^5} \right)^{-n/2(n+1)} \,.
\end{equation}

In polytropes, the sound speed is proportional to $x^{1/2}$
\begin{equation} \label{eq:c_s_x}
    c_s(x) = \frac{1}{\sqrt{n}} \sqrt{\frac{GM_\star}{R_\star}} \left( \frac{x}{R_\star} \right)^{1/2} \,,
\end{equation}
or in terms of density
\begin{equation} \label{eq:c_s_rho}
    c_s(\rho) = K_{c_s}(n) \sqrt{\frac{GM_\star}{R_\star}} \left(\frac{\rho}{M_\star/R_\star^3} \right)^{1/2n} \,.
\end{equation}
By setting $v_{\rm sh}(\rho_s) = c_s(\rho_s)$ we find the density at which the shock energy has been deposited, {using the GFKS evolution, $v_{\rm sh} \propto \rho^{-\mu}$ (equation \ref{eq:v_sh_rho})}
\begin{equation}
    \rho_{s} = K_{\rho_s}(n) \frac{M_\star}{R_\star^3} \left( \frac{G \kappa^2 M_\star^3}{c^2 R_\star^5} \right)^{-\frac{n^2 \mu}{(n+1)(1+2n\mu)}} \,,
\end{equation}
and finally, from equation \ref{eq:E_dep_min} we get the minimal energy
\begin{equation}
    E_{\rm dep,min} = K_{E}(n) \frac{G M_\star^2}{R_\star} \left( \frac{G \kappa^2 M_\star^3}{c^2 R_\star^5} \right)^{-\frac{n\mu (n+2)}{(1+2n\mu)(n+1)}} \,,
\end{equation}
or when scaled by the polytrope's gravitational potential energy
\begin{equation}
    \frac{E_{\rm dep,min}}{|E_{\rm bind}|} = K_{\rm bind}(n) \left( \frac{G \kappa^2 M_\star^3}{c^2 R_\star^5} \right)^{-\frac{n\mu (n+2)}{(1+2n\mu)(n+1)}} \,.
\end{equation}

For a compact progenitor with radiative envelope ($n=3$) we get a minimal energy of
\begin{equation}
    E_{\rm dep,min} = 5.8 \times 10^{44} \,
    M_{10}^{1.0} \, R_{2}^{0.66} \, \kappa_{0.3}^{-0.66} \, \rm erg \,,
\end{equation}
and correspondingly, a minimal ejecta mass of
\begin{equation} \label{eq:m_bo_compact}
    m_{\rm bo} = 5.9 \times 10^{-8} \, \, M_{10}^{-1/2} \, R_{2}^{5/2} \, \kappa_{0.3}^{-1} \, \, \rm M_\odot \,,
\end{equation}
following equation \ref{eq:m_bo}.

For a red supergiant, modelled by an $n=3/2$ polytrope, we find
\begin{equation}
    E_{\rm dep,min} = 3.0 \times 10^{46} \,
    M_{10}^{1.16} \, R_{500}^{0.40} \, \kappa_{0.3}^{-0.56} \, \rm erg \,,
\end{equation}
and minimal ejecta mass
\begin{equation} \label{eq:m_bo_extended}
    m_{\rm bo} = 5.8 \times 10^{-2} \, \, M_{10}^{-1/2} \, R_{500}^{5/2} \, \kappa_{0.3}^{-1} \, \rm M_\odot \,.
\end{equation}

\begin{table}
\caption{Dimensionless coefficients in the analytical expression obtained for polytropes.}
\begin{tabular}{|c c c c|}
\hline
{\bf Constant}	& Expression & {$n=3/2$}	& {$n=3$} \\ \hline

$K_\rho(n)$ & -  & 0.92 & 0.32 \\ \hline

$K_\tau(n)$ & $\left( (n+1) K_\rho^{1/n} \right)^{-1}$ & 0.42 & 0.36 \\ \hline

$K_{\rho,bo}(n)$ & $\left( \frac{C_{\rm f}}{3\sqrt{2}K_\tau}\right)^{n/(n+1)}$ & 1.07 & 1.21 \\ \hline

$K_{c_s}(n)$ & $\left( n K_\rho^{1/n} \right)^{-1/2}$ & 0.84 & 0.70 \\ \hline

$K_{\rho_s}(n)$ & $\left( \frac{K_{\rho,bo}^\mu \sqrt{2}}{K_{c_s} C_{\rm f}} \right)^{2n/(2n\mu+1)}$ & 0.75 & 1.16 \\ \hline

$K_{E}(n)$ & $\frac{4\pi}{n(n+1)} K_{\rho_s}^{1+2/n} K_\rho^{-2/n}$ & 1.93 & 2.83 \\ \hline

$K_{\rm bind}(n)$ & $K_E \frac{5-n}{3}$ & 2.25 & 1.88
\\ \hline
  
\end{tabular}
\label{tab:Polytropes_Constants}
\end{table}%

\section{Shocks launched from the stellar core}\label{sec:Shocks_from_core}

So far, we have discussed mass ejection by examining the conditions at the breakout layer, and backtracking inwards (and earlier) into the star, in order to infer the minimal energy required for mass ejection. In this section, we take the complementary approach by considering an example where energy is deposited at the center of a star, and analyzing the resulting mass loss. We demonstrate how the lower limits we previously found are indeed satisfied.

\subsection{Analytical estimates} \label{sec:SmallExp_Analytical}
Consider a point explosion of energy $E$ set at the center of a star, with $E \ll E_{\rm bind}$. As before, we assume that the star is a polytrope and derive approximate analytical expressions. In the following sections we discuss the 4 phases of the shock propagation through the star - Sedov-Taylor phase, spherical acoustic propagation, planar weak shock, and lastly planar strong shock.

The propagation and outcomes of point explosions set at the center of a star have been extensively studied, both analytically and numerically, in the context of supernovae explosions \citep[e.g.,][]{Matzner_McKee_1999}. However, here we are interested in \textit{small} explosions, where the deposited energy is clearly insufficient for unbinding the majority of the of the stellar mass. 

\subsubsection{Sedov-Taylor Point explosion}
The famous point explosion problem was originally solved independently by Taylor, von-Neumann and Sedov, using a self-similarity argument \citep{Taylor_1950,Bethe_1958,Sedov_1959}. Conservation of energy gives the scaling of the shock radius with time, while the pressure, density and velocity profiles within the shocked region can be found analytically through the self-similar ansatz.

Formally, self-similarity is obtained just when the ambient density profile is scale free (i.e., uniform, or scaling as a power-law of distance), while the density in a polytrope varies on a scale $R_\star$. However, at sufficiently small radii, $r \ll R_\star$ the density is uniform up to first order in $r$, thanks to the inner boundary condition $d\rho/dr|_{r=0}=0$. 

The Sedov-Taylor solution also assumes that the shock is strong - i.e., ambient pressure is negligible with respect to the post-shock pressure. This assumption is initially satisfied, but as the blast wave expands, the post-shock pressure decreases as $p \propto r^{-3}$, and the shock becomes decreasingly strong.

\subsubsection{Spherical acoustic expansion}
When the pressure behind the shock becomes comparable to the ambient counter pressure, the Sedov-Taylor solution no longer holds. {Neglecting dimensionless prefactors, this condition can be expressed as}
\begin{equation}
    \rho_c v_{\rm sh}^2 \approx p_c \approx \frac{GM_\star^2}{R_\star^4} \,,
\end{equation}
{where $\rho_c$ and $p_c$ are the central density and pressure, respectively. The shocked region of a Sedov-Taylor explosion in uniform density remains at causal contact, and thus the post-shock pressure is comparable to average energy density, $\rho_c v_{sh}^2 \approx E/R^3$.} Thus, the transition radius is of order
\begin{equation}
    R_{ST} \sim R_\star \left( \frac{E}{E_{\rm bind}} \right)^{1/3} \,,
\end{equation}
where we assumed that the ambient density and pressure are roughly constant up to $R_{ST}$. 
Note that for the energies we are interested in, where $E\ll E_{\rm bind}$, this transition radius is well below the stellar radius, where, for polytropic stars, the density is roughly constant.

This scale also marks the transition from a strong shock to a spherically expanding sound pulse. The pulse propagates at the local sound speed, which is similar to the central sound speed, up to a radius of about $r\approx R_\star/2$. The material velocity at $R_{ST}$ is, by definition, of the same order of the central sound speed, $\delta v(R_{ST}) \approx c_s(R_{ST}) \approx v_{\rm esc}$. 

In the absence of dissipation, the amplitude of a spherically expanding linear sound pulse decays as $1/r$, while its width remains constant, such that wave energy is conserved. In principle, non-linearity results in steepening and distortion of the wave profile into a weak shock, that dissipates energy in the form of heat as it propagates. However, as discussed in \cite{Landau_Lifshitz_1987}, Section 102, even after shock formation, the pulse width increases very slowly with distance, scaling as $\sqrt{\log{r}}$, while the amplitude decreases only slightly faster than a linear sound pulse, scaling as $1/(r \sqrt{\log{r})}$.

Neglecting these logarithmic corrections, the material velocity at the end of the spherical expansion phase, i.e., at $r\approx R_\star/2$ is roughly
\begin{equation}
    \delta v(r\approx R_\star/2) \approx v_{\rm esc} \left( \frac{R_{ST}}{R_\star} \right) = v_{\rm esc} \left( \frac{E}{E_{\rm bind}} \right)^{1/3} \,.
\end{equation}

Note that this regime is obtained only when $E\ll E_{\rm bind}$, such that $R_{ST} \ll R_\star$. Otherwise, the shock remains strong through the entire star, and never transitions to a sound wave.

\subsubsection{Weak planar shock} \label{sec:weak_planar_shock}
At the outer half of the star, the radial coordinate increases only by a factor of 2, while temperature and density vary from roughly their central value, to zero at the stellar surface. To a rough approximation, at the range $(R_\star/2,R_\star)$ the {radial coordinate} can be taken as constant, and the propagation is {nearly plane-parallel, neglecting the geometrical effects of the spherical symmetry}.

Two opposite dispersive effects take place during this phase. The pulse's finite width implies that the medium's sound speed varies across the pulse as
\begin{equation}
    \frac{\Delta c_s}{c_s(x)} \approx \frac{1}{2} \frac{\Delta x}{x} \,,
\end{equation}
where $x$ is the pulse's distance from the surface, and $\Delta x$ is its width (equation \ref{eq:c_s_x}). Since the pulse's leading (outer) edge is always positioned at a region colder than its trailing edge, this effect tends to decrease the pulse's width as $\Delta \dot{ x }_{\rm width} \approx -\Delta c_s = -c_s(x) \Delta x / (2x) < 0$.

The second dispersive effect arises from the flow equations' non-linearity. Variations in material velocity imply that the pulse's peak propagates faster than its node, thus widening at a rate $\Delta \dot{x}_{nl} \approx \frac{\gamma+1}{2} \delta v$, where $\delta v$ is the amplitude of the material velocity across the pulse. Here we used the fact that adiabatic perturbations in velocity are related to perturbations in the local sound speed, as $\delta(c_s) = \frac{\gamma-1}{2} \delta v$, where $\gamma$ is the adiabatic index of the gas.

Which of the two effects - compression due to finite width, versus non-linear widening, is dominant in shaping the pulse's width? We define
\begin{equation}
    Z = \left| \frac{\Delta \dot{x}_{nl}}{\Delta \dot{x}_{\rm width}} \right| \approx \frac{\delta v / c_s}{\Delta x / x} \,,
\end{equation}
a dimensionless number comparing the two effects. Non-linear widening dominates when $Z \gg 1$, and compression is dominant when $Z \ll 1$. In appendix \ref{sec:app_weak_planar_shock} we show that the two effects tend towards equilibrium - if initially $Z \ll 1$, it increases as the pulse approaches the surface, and vice versa - $Z$ decreases as long as non-linearity dominates ($Z \gg 1$). Inevitably, any weakly non-linear pulse evolves towards a state with $Z \sim 1$. Remarkably, at the onset of the planar phase, the two dispersive effects are similarly important
\begin{equation} \label{eq:eta_weak_planar}
    \left. \frac{\Delta x}{x} \right|_{r\approx R_\star/2} \approx \left. \frac{\delta v}{c_s} \right|_{r\approx R_\star/2} \approx \left( \frac{E}{E_{\rm bind}} \right)^{1/3} \,,
\end{equation}
such that initially $Z \sim 1$, and we therefore conclude that the pulse's amplitude and width evolve jointly, such that the ratio of the Mach number and the relative width is conserved.

Our goal is to solve for the evolution of $\Delta x$ and $\delta v$ during the weak planar shock phase, remembering that dissipation can modify the shock properties such that energy conservation can no longer be assumed. Approximating their evolution as a power-law of $x$, we obtain
\begin{equation} \label{eq:v_weak_planar}
    \delta v \propto x^{-\varepsilon_{\eta}+1/2} \,,
\end{equation}
and
\begin{equation}
    \Delta x \propto x^{1-\varepsilon_{\eta}} \,,
\end{equation}
such that $\Delta x/x \sim \delta v / c_s \propto x^{-\varepsilon_\eta}$, and $Z = \rm const$. The pulse's energy scales as
\begin{equation} \label{eq:E_weak_planar}
    E \approx \rho r^2 \, \Delta x \, \delta v^2 \approx x^{n+2 - 3\varepsilon_\eta} \,,
\end{equation}
where $r \approx \rm const$ in this regime.

The material velocity must increase as the pulse approaches the surface, therefore $\varepsilon_\eta > 1/2$ from equation \ref{eq:v_weak_planar}. Additionally, the pulse energy can only decrease as it propagates and $x$ decreases, so equation 
\ref{eq:E_weak_planar} entails $\varepsilon_\eta < (n+2)/3$. Together, these constraints are
\begin{equation} \label{eq:epsilon_eta_range}
     1/2 < \varepsilon_\eta < (n+2)/3 \,.
\end{equation}

Since $\varepsilon_\eta > 0$, the shock becomes increasingly strong as it approaches the stellar surface. At some distance, $x_s$ from the surface the material velocity becomes sonic, defined by
\begin{equation}
    \frac{\delta v}{c_s} \approx \left( \frac{E}{E_{\rm bind}} \right)^{1/3} \left( \frac{x_s}{R_\star} \right)^{-\varepsilon_\eta} = 1 \,,
\end{equation}
where we used the value at $R_\star/2$, at the onset of this phase (equation \ref{eq:eta_weak_planar}), and the scaling $\delta v/c_s \propto x^{-\varepsilon_\eta}$. Therefore, the depth $x_s$ of strong shock formation is
\begin{equation} \label{eq:x_s_E}
    \frac{x_s}{R_\star} \approx \left( \frac{E}{E_{\rm bind}} \right)^{1/(3\varepsilon_\eta)} \,,
\end{equation}
at which point the material velocity is roughly (see equation \ref{eq:c_s_x})
\begin{equation}
    \delta v(x_s) = c_s (x_s) \approx v_{\rm esc} \left( \frac{E}{E_{\rm bind}} \right)^{1/(6\varepsilon_\eta)} \,.
\end{equation}

\subsubsection{Strong planar shock}
Lastly, at $x<x_s$ the pulse's front is a strong shock, propagating in a power-law density profile, and it is therefore described by the GFKS self-similarity solution. The flow's self-similarity implies that its characteristic width is always similar to the shock's distance from the surface. We note that the fact that $\Delta x / x \sim \delta v/c_s$ throughout the weak-shock planar phase implies that $\Delta x(x_s) \approx x_s$, allowing the flow to smoothly transition to the GFKS solution at $x_s$. Note that when $E \gg E_{\rm bind}$ (the case in supernovae explosions, for example), the Sedov-Taylor phase transitions directly to the planar Sakurai phase, as is showcased by equation 19 of \cite{Matzner_McKee_1999}.

In the problem considered here, we use the result of \cite{Sakurai_1960}, $\delta v \propto x^{-\mu n}$, to find
\begin{equation}
    \delta v = v_{\rm esc} \left( \frac{E}{E_{\rm bind}} \right)^{1/6\varepsilon_\eta} \left( \frac{x}{x_s} \right)^{-n \mu} \,.
\end{equation}
The material velocity approaches the escape velocity at a depth $x_{\rm esc}$, given by
\begin{equation}
    \frac{x_{\rm esc}}{R_\star} = \left( \frac{E}{E_{\rm bind}} \right)^{(2n\mu+1)/6n\mu \varepsilon_\eta} \,,
\end{equation}
where equation \ref{eq:x_s_E} was used, and hence the mass ejected by the shock scales as
\begin{equation} \label{eq:m_E_theoretical}
    \frac{m_{\rm ej}}{M_\star} \approx \left( \frac{x_{\rm esc}}{R_\star} \right)^{n+1} = \left( \frac{E}{E_{\rm bind}} \right)^{(2n\mu+1)(n+1)/6n\mu \varepsilon_\eta} \,.
\end{equation}

Denoting $m_{\rm ej} \propto E^{\varepsilon_m}$, we find from equation \ref{eq:epsilon_eta_range} the following possible range for $\varepsilon_m$
\begin{equation} \label{eq:epsilon_m_range}
    \frac{(2n\mu+1)(n+1)}{2n\mu (n+2)} < \varepsilon_{m} < \frac{(2n\mu +1)(n+1)}{3n\mu} \,,
\end{equation}
where the left inequality corresponds to energy conservation in the weak shock planar phase. For $n=3$, $\mu = 0.188$ we find
\begin{equation} \label{eq:epsilon_m_range_n=3}
    1.51 < \varepsilon_m < 5.03 \,,
\end{equation}
and for $n=3/2$, $\mu = 0.222$
\begin{equation} \label{eq:epsilon_m_range_n=3/2}
    1.79 < \varepsilon_m < 4.17 \,.
\end{equation}
Hence, the ejected mass will be a strong function of shock energy. 

Note that for any $n>0$, $\varepsilon_m > 1$. The limit $\varepsilon_m = 1$ is achieved only if the energy $E$ is invested in its entirety in accelerating the ejected material to the escape velocity, i.e., when $E \approx m_{\rm ej}v_{\rm esc}^2$. In the process described here, some of the deposited energy is spent on heating deep layers that are not being ejected. The fact that $\varepsilon_m>1$ also implies that for a given amount of energy, mass ejection is maximized by a single explosion, rather than a sequence of smaller explosions of the same total energy.

\section{Numerical investigations} \label{sec:numerical_investigations}

To verify our analytical results and to determine the values of $\varepsilon_\eta$ and $\varepsilon_m$ that characterize shock evolution and mass ejection from weak shocks, we turn to numerical investigations. We simulate the point explosion problem using the 1D version of the hydrodynamical code RICH \citep{Yalinewich_2015}. We construct a polytrope, characterized by an equation of state $p=s\rho^{1+1/n}$, where $s$ is a constant. A small "hotspot" region is set at the center of the polytrope, with radius $R_{\rm HS}$ and pressure $p_{\rm HS}$. The extra energy in the explosion is given by
\begin{equation}
    \frac{E}{E_{\rm bind}} = A_E \left( \frac{p_{\rm HS}}{p_c} \right) \left( \frac{R_{\rm HS}}{R_\star} \right)^3 \,.
\end{equation}
where $p_c$ is the polytrope's central pressure, $p_c=s\rho_c^{1+1/n}$. $A_E$ is a dimensionless coefficient given by
\begin{equation}
    A_E = \frac{4\pi n(5-n)}{9} \frac{p_c R_\star^4}{G M_\star^2} \,,
\end{equation}
which is $A_E=5.6$ for $n=3/2$ and $A_E=92.6$ for $n=3$. We fix $R_{\rm HS} = 10^{-2}$, and vary $p_{\rm HS} = \{10,10^{1.5},10^2,10^{2.5},10^3,10^{3.5}\}$.

We are interested in calculating the amount of mass ejected by the accelerating shock wave. First we find the depth $x_{\rm esc}(E;n)$, defined as the shock position at which the peak material velocity, just behind the shock is
\begin{equation}
    \delta v (x_{\rm esc}) =  \frac{2}{(\gamma+1)C_{\rm f}} v_{\rm esc} \,,
\end{equation}
where the prefactor accounts for the velocity in the immediate downstream of a strong shock, and the acceleration factor $C_{\rm f} \approx 2$ long after shock passage. All matter exterior to $x_{\rm esc}$ eventually accelerates to beyond $v_{\rm esc}$ and becomes unbound. The ejected mass is given by
\begin{equation}
    m_{\rm ej} = \int_{R_\star-x_{\rm esc}}^{R_\star} 4\pi r^2 \rho(r) \, dr \approx \frac{4\pi R_\star^2}{(n+1)} \rho(x_{\rm esc}) x_{\rm esc} \,,
\end{equation}
where we used equation \ref{eq:rho_x} in the last approximation, valid when $x_{\rm esc} \ll R_\star$.

\begin{figure}
    \centering
    \includegraphics[width=\columnwidth]{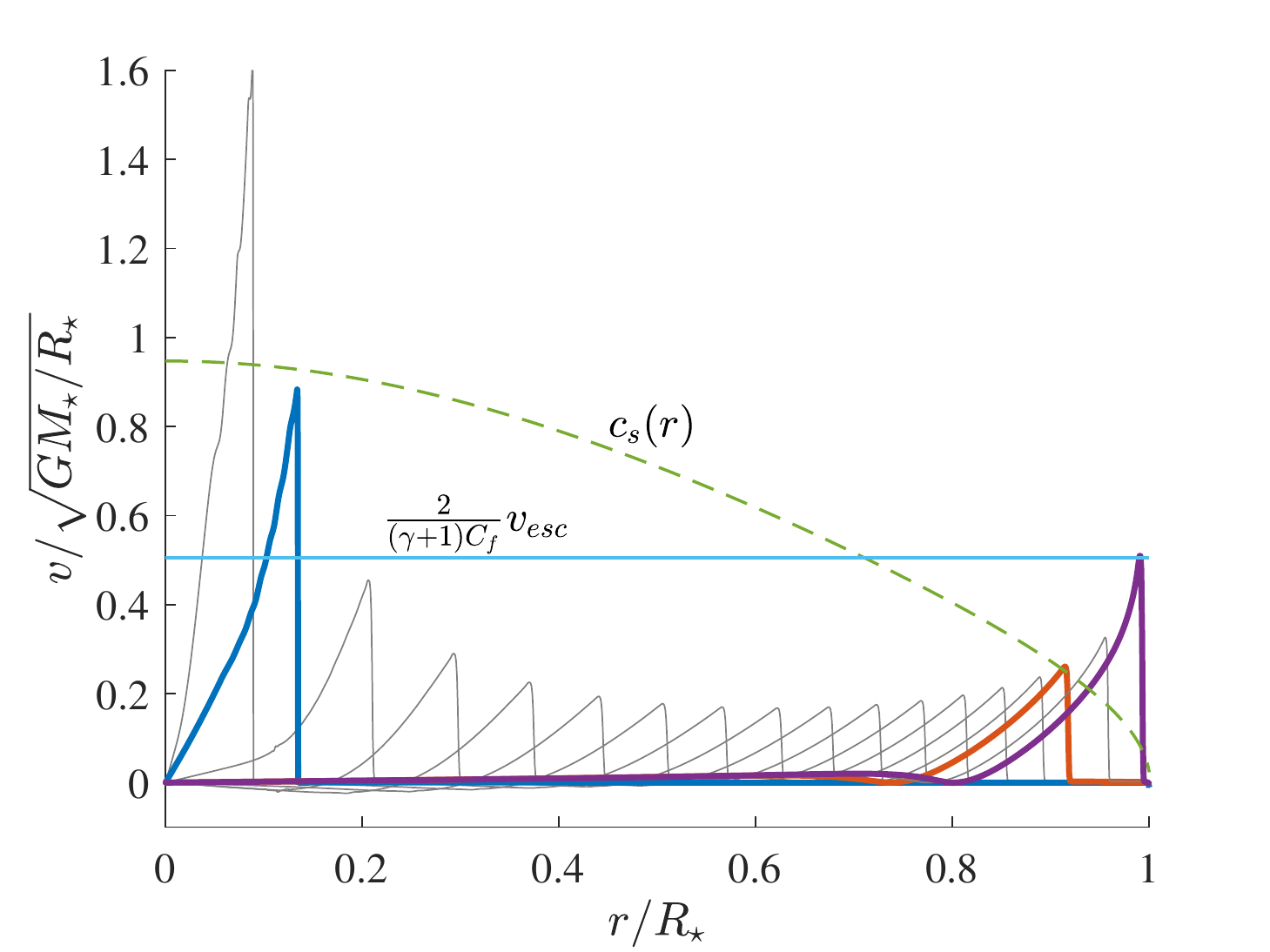}
    \caption{The material velocity profile at different times after an explosion with energy $E/E_{\rm bind}=0.018$ set at the center of an $n=3/2$ polytrope.  The \textit{dashed green line} is the speed of sound in the undisturbed star, and the \textit{horizontal solid blue line} is the threshold post-shock velocity that results in mass ejection, as described in the text. Highlighted in \textit{blue} is the velocity profile at the end of the Sedov-Taylor phase, when the shock first becomes weak. The pulse is then subsonic throughout most of the star, until it transitions to a strong shock, shown in \textit{red}, when the material velocity becomes sonic. Finally, in \textit{purple} we show the velocity profile when the ejection criterion is first met, at depth $x_{\rm esc}\approx 10^{-2} R_\star$ measured from the stellar edge.}
    \label{fig:ExampleVelocityProfile}
\end{figure}

Figure \ref{fig:ExampleVelocityProfile} shows the material velocity profile as the shock propagates through the star, for an explosion energy of $E/E_{\rm bind} = 0.018$. This figure demonstrates the definition of $x_s$ - the depth at which material velocity first becomes sonic (red curve), and $x_{\rm esc}$ - where the post-shock material velocity is eventually accelerated to $v_{\rm esc}$ (purple curve).

\begin{figure}
    \centering
    \includegraphics[width=\columnwidth]{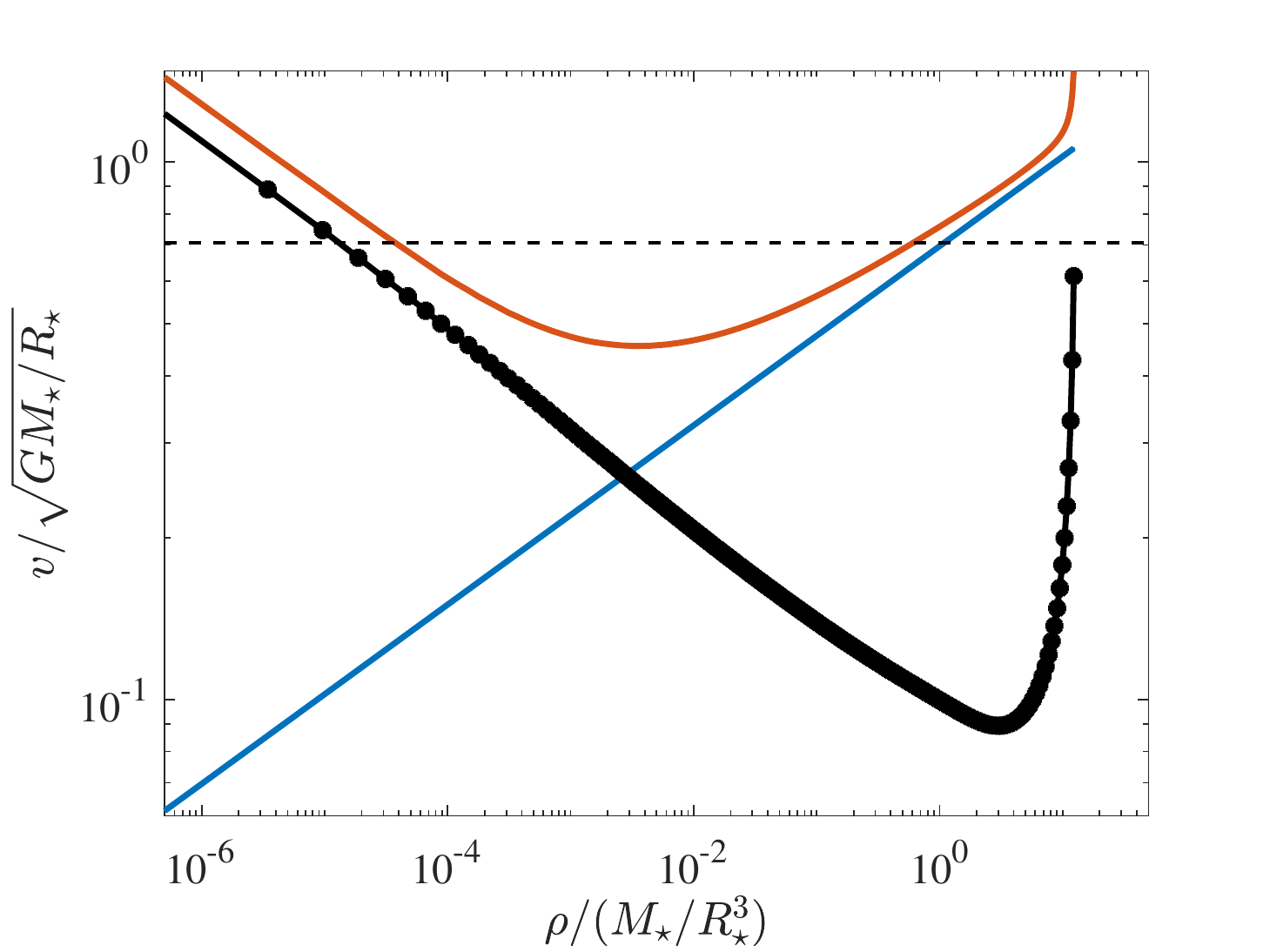}
    \caption{Shock and material velocity as a function of unperturbed density, for a simulation of an explosion of energy $E/E_{\rm bind} = 2.9\times 10^{-3}$ set in an $n=3$ polytrope. The numerical results in this figure mirror the analytic picture in figure \ref{fig:VisualizeMinShockEnergy}. Peak material velocity is plotted in \textit{black}, and the unperturbed sound speed in \textit{blue}. The shock velocity (in \textit{red}) was calculated from the material velocity and the local Mach number with respect to the material lying just ahead of the shock. Finally, the \textit{horizontal dashed line} is the threshold shock velocity for mass ejection, $v_{\rm esc}/C_{\rm f}$. }
    \label{fig:NumericalVelocity_vs_Density}
\end{figure}

Figure \ref{fig:NumericalVelocity_vs_Density} provides a complementary view on the results of the numerical simulations, showing the velocity evolution throughout the stellar interior, demonstrating the analytical picture portayed in figure \ref{fig:VisualizeMinShockEnergy}. During the spherical phases, the shock and material velocities decrease while density remains roughly the same. Later on, the pulse continues to decelerate since the ambient temperature drops in the outer layers of the star, while the (subsonic) material velocity behind the shock increases. At some point the material velocities approach the local sound speed, and the shock begins to accelerate, transitioning to a strong shock. The material outside the layer at which $v_{\rm sh}=v_{\rm esc}/C_{\rm f}$ is being ejected by the explosion. 

\begin{figure}
    \centering
    \includegraphics[width=\columnwidth]{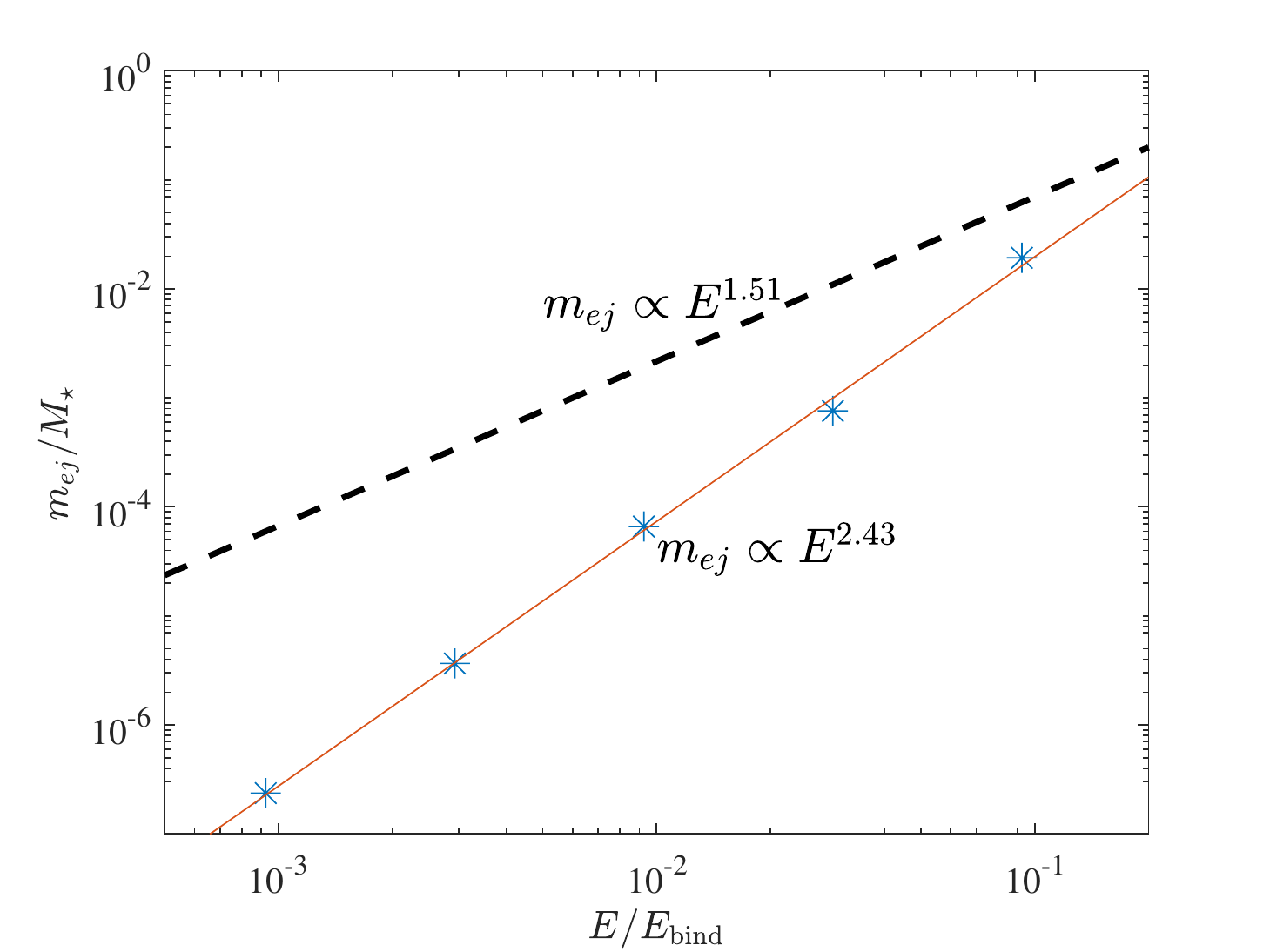}
    \caption{Mass ejection from a point explosion in an $n=3$ polytrope. \textit{Stars} are the numerical results and the \textit{solid red line} is a fit to a power-law. The mass ejection efficiency is consistently lower than the theoretical upper limit, see equation \ref{eq:m_E_theoretical} (\textit{dashed black line}).}
    \label{fig:NumericalMassEnergy_n=3}
\end{figure}

\begin{figure}
    \centering
    \includegraphics[width=\columnwidth]{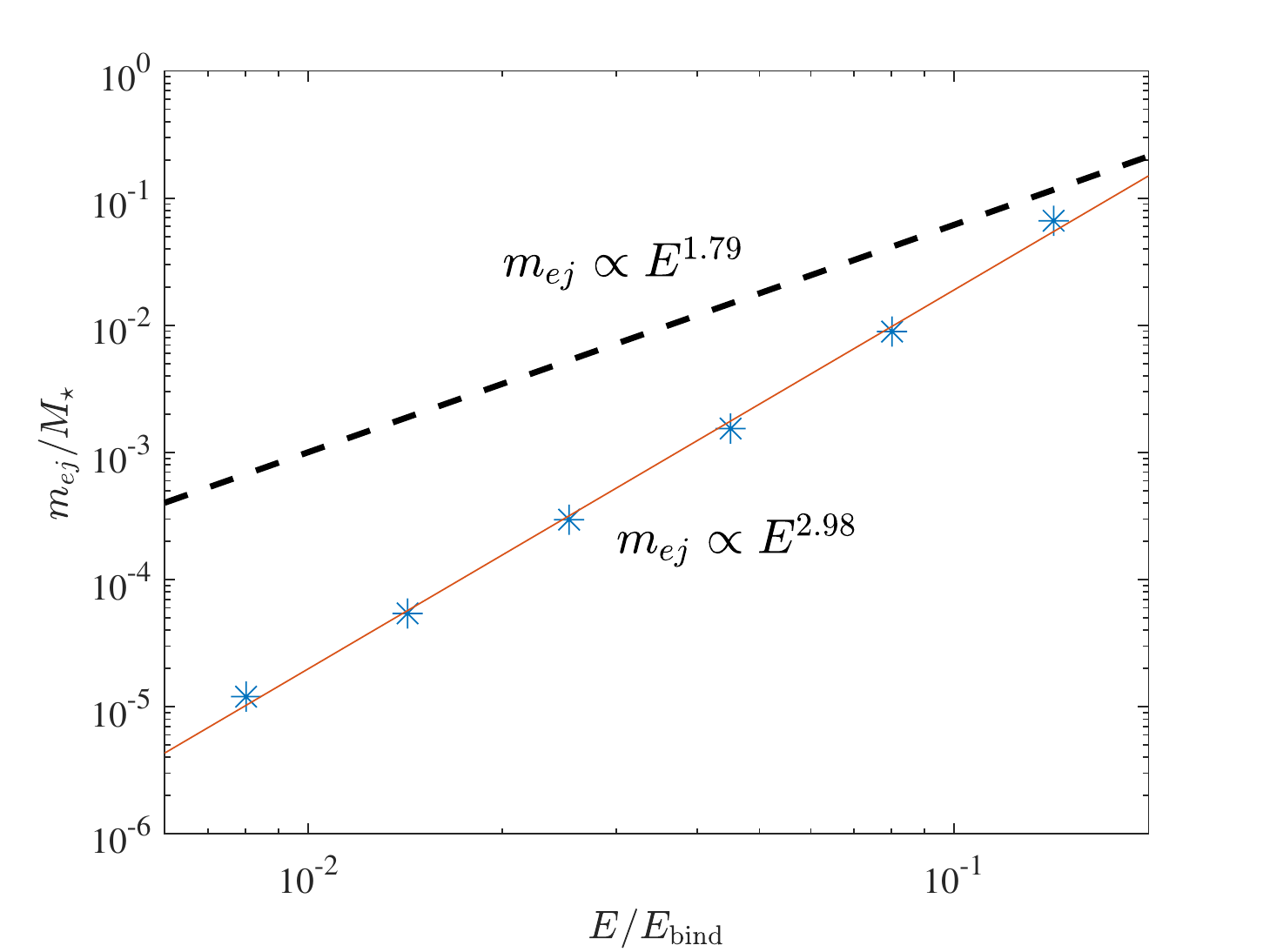}
    \caption{Same as figure \ref{fig:NumericalMassEnergy_n=3}, for $n=3/2$. Here the theoretical upper limit (\textit{dashed black line}) is given in equation \ref{eq:epsilon_m_range_n=3/2}.}
    \label{fig:NumericalMassEnergy_n=1_5}
\end{figure}

Our results are presented in figures \ref{fig:NumericalMassEnergy_n=3} and \ref{fig:NumericalMassEnergy_n=1_5}. We show the numerical results from a set of simulations, with a fit to a power-law trend. Alongside, we show the analytical upper limit on the mass ejection discussed in section \ref{sec:SmallExp_Analytical}, equations \ref{eq:epsilon_m_range_n=3} and \ref{eq:epsilon_m_range_n=3/2}. As predicted, the mass yield that small point explosions produce is smaller than the theoretical upper limit. The fit to the numerical results is of the form
\begin{equation}
    \frac{m_{\rm ej}}{M_\star} = B_m(n) \, \left( \frac{E}{E_{\rm bind}} \right)^{\varepsilon_m(n)} \,,
\end{equation}
with $B_m = 5.28$, $\varepsilon_m = 2.43$ for $n=3$ polytropes, and $B_m = 18.14$, $\varepsilon_m = 2.98$ for $n=3/2$. 

Taking the inferred value of $\varepsilon_m$ we use equation \ref{eq:m_E_theoretical} to obtain $\varepsilon_\eta = 1.04$ for $n=3$ and $\varepsilon_\eta = 0.70$ for $n=3/2$. 

As discussed in section \ref{sec:min_energy_general}, accelerating radiative shocks cannot unbind less than the breakout mass, given in equation \ref{eq:m_bo}. We therefore conclude that the minimal explosion energy that results in any mass ejection is
\begin{equation}
    \frac{E}{E_{\rm bind}} > \left( \frac{4\pi C_{\rm f}}{3\sqrt{2}B_m(n)} \right)^{1/\varepsilon_m(n)} \left( \frac{G \kappa^2 M_\star^3}{c^2 R_\star^5} \right)^{-1/2\varepsilon_m(n)} \,.
\end{equation}
Scaled to typical values, we find for $n=3$
\begin{equation} \label{eq:E_min_numerical_n=3}
    E > 5.89 \times10^{46} \, M_{10}^{1.38} \kappa_{0.3}^{-0.41} R_2^{0.03} \, \rm erg \,,
\end{equation}
and for $n=3/2$
\begin{equation} \label{eq:E_min_numerical_n=1.5}
    E > 4.38 \times 10^{46} \, M_{10}^{1.50} \kappa_{0.3}^{-0.33} R_{500}^{-0.16} \, \rm erg \,.
\end{equation}

Note that the minimal energy depends weakly on stellar radius (especially for $n=3$ polytropes). Additionally, the minimal energy is of the same order of magnitude for both progenitor types considered here. However, while the limiting energy is similar, the corresponding minimal ejecta mass strongly depends on the stellar radius, varying by many orders of magnitude across different progenitors (equations \ref{eq:m_bo_compact} and \ref{eq:m_bo_extended}). 

\begin{figure}
    \centering
    \includegraphics[width=\columnwidth]{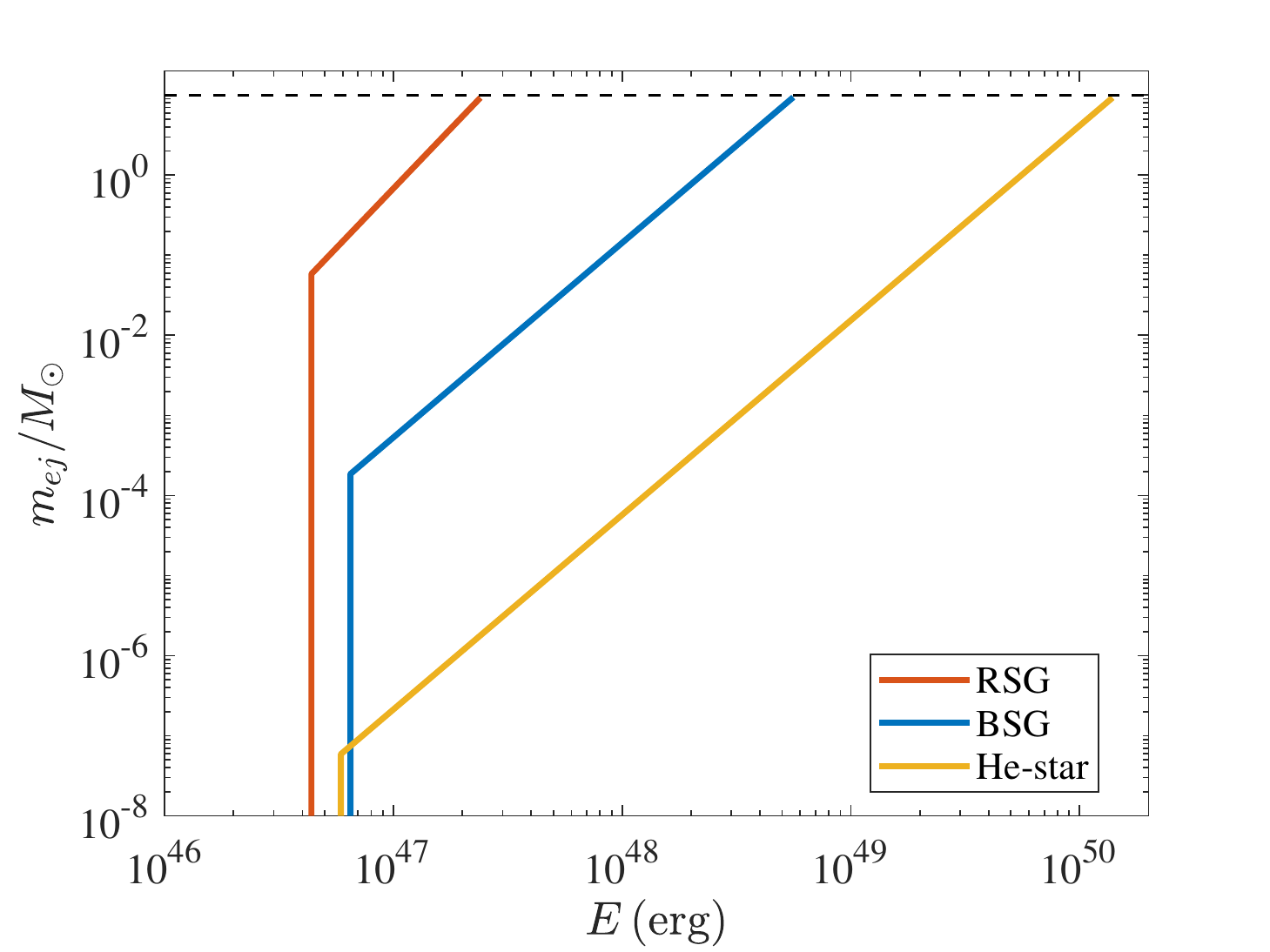}
    \caption{Mass ejection as a function of explosion energy, for different progenitors. All three stars were modelled as polytropes with a mass of $\rm 10 \, M_\odot$. The red supergiant (RSG) has a radius of $500 \, \rm R_\odot$ and $n=3/2$, while the blue superigant (BSG) and helium star have radii of $50 \, \rm R_\odot$ and $2 \, \rm R_\odot$ respectively, both modelled with $n=3$. The \textit{horizontal dashed line} corresponds to the entire stellar mass. All curves terminate at some minimal explosion energy, below which no mass escapes the star (equations \ref{eq:E_min_numerical_n=3} and \ref{eq:E_min_numerical_n=1.5}). }
    \label{fig:MassEnergy_progenitors}
\end{figure}

We summarize these results in figure \ref{fig:MassEnergy_progenitors}, showing the ejecta mass as a function of energy for a few characteristic stars. Red supergiants have a fairly small energy range for which partial mass ejection is achieved - less than one order of magnitude in energy. This implies that these stars will either lose no mass, or they are fully disrupted, unless the explosion energy is finely tuned.

More compact progenitors, like blue-supergiants or Helium stars have larger binding energies, and thus stronger explosions are required to completely disintegrate them. However, since their envelopes are optically thicker than those of RSG's, radiation mediated shocks break out closer to the stellar surface, and can thus accelerate small amounts of mass to beyond the star's escape velocity. Explosions with $E \ll E_{\rm bind}$ can therefore produce mass ejection across a wide energy range in compact progenitors (almost 3 orders of magnitude for Helium stars).

\section{Radiative shocks}
In our derivation of the minimal energy required for mass ejection, we assumed that the shock is radiation mediated (e.g., equations \ref{eq:tau_bo} and \ref{eq:tau_bo_limit}), such that it stops accelerating as it approaches the breakout depth. Here we examine the validity of this assumption in different progenitor stars.

A shock is \textit{radiation dominated} if the downstream pressure is dominated by photon rather than gas pressure. A strong shock propagating into a medium of density $\rho$ with velocity $v_{\rm sh}$ is therefore radiation dominated if the ratio
\begin{equation} \label{eq:U_ratio}
    \frac{U_{\rm rad}}{U_g} = \frac{8(\gamma-1)^5}{(\gamma+1)^7} \frac{a m^4}{k_B^4} \frac{v_{\rm sh}^6}{\rho} \,,
\end{equation}
is much greater than 1, where $\gamma$ is the downstream's effective adiabatic index, $a$ is the radiation constant, $m$ is the mean particle mass, and $k_B$ the Boltzmann constant. Here we assumed that photons are in thermal equilibrium with the gas, which is valid for non-relativistic shocks, with $v_{\rm sh} \lesssim 0.1 \, c$ \citep{Nakar_Sari_2010}. 

We are interested in the shock properties as it approaches the breakout layer. In the limiting case, the breakout density is given by equation \ref{eq:rho_bo}, and the shock's velocity is $v_{\rm esc}/C_{\rm f}$. Plugging in equation \ref{eq:U_ratio} we get
\begin{equation}
    \frac{U_{\rm rad}}{U_g} = \frac{2^6 (\gamma-1)^5}{C_{\rm f}^6 (\gamma+1)^7 K_{\rho,bo}} \frac{a G^3 m^4}{k_B^4} M_\star^2 \left( \frac{G \kappa^2 M_\star^3}{c^2 R_\star^5} \right)^{n/2(n+1)} \,.
\end{equation}

For $n=3$, $\gamma = 4/3$, we get the following energy density ratio
\begin{equation} \label{eq:U_ratio_n=3}
    \left. \frac{U_{\rm rad}}{U_g} \right|_{n=3} \approx 970 \: M_{10}^{25/8} R_2^{-15/8} \kappa_{0.3}^{3/4} \left( \frac{m}{m_p} \right)^4 \,,
\end{equation}
where $m_p$ is the proton mass. For $n=3/2$, $\gamma = 5/3$, we find
\begin{equation} \label{eq:U_ratio_n=3/2}
    \left. \frac{U_{\rm rad}}{U_g} \right|_{n=3/2} \approx 0.16 \: M_{10}^{29/10} R_{500}^{-3/2} \kappa_{0.3}^{3/5} \left( \frac{m}{m_p} \right)^4 \,.
\end{equation}

Equation \ref{eq:U_ratio_n=3} implies that the downstream of marginally mass-ejecting shocks is radiation dominated in compact progenitors with radiative envelopes (such that $n=3$). This also justifies the use of $\gamma=4/3$ for the shocked downstream in this regime, which is a good approximation when the shock is radiation dominated. Our calculation of the minimal ejecta mass in section \ref{sec:min_energy_general} is thus entirely self-consistent - the shock front spans an optical depth $c/(3v_{\rm sh})$, and our assumptions are valid.

The case of extended progenitors, described by equation \ref{eq:U_ratio_n=3/2} is more subtle. The slower escape velocity of these stars significantly reduces the importance of radiation pressure in the shocked downstream, due to the strong dependence on $v_{\rm sh}$ in equation \ref{eq:U_ratio}. However, even when the downstream pressure is not dominated by radiation, photons still play an important role in shaping the shock front, as long as the photon energy flux is important relative to the advective flux of gas pressure, i.e., when $c U_{\rm rad} > v_{\rm sh} U_g$ \citep[e.g.,][Chapter VII, Section 14]{Zeldovich_Raizer_1967}. The shock is then preceded by a radiative precursor, heating the upstream ahead of the shock. The width of the heated region then sets the distance from the stellar edge at which shock acceleration is terminated.

Accurately solving the shock structure in the regime $v_{\rm sh}/c \ll U_{\rm rad}/U_g \lesssim 1$ is beyond the scope of this work. We note however that when opacity is dominated by absorption/emission, applicable in the atmospheres of red supergiants, where H- is the dominant source of opacity, the derivation presented in \citealt[][Chapter VII, Section 17]{Zeldovich_Raizer_1967}, applies, and the resulting shock front spans an optical depth of approximately $cU_{\rm rad}/(v_{\rm sh} U_g)$, yielding different scaling relations than those derived in this work. We leave the treatment of the problem in the case of very extended progenitors to a future study, where non-uniform opacity at the outer stellar layers will be considered. 



\section{Discussion and Conclusions} \label{sec:discussion}

Our work has important applications to low-energy explosions from massive stars, such as luminous blue variables, pre-supernova outbursts, and failed supernovae. Assuming such outbursts result from energy deposition well below the photosphere, our results place a lower limit on the energy budget of the underlying mechanism. Remarkably, across a wide range of progenitor radii, outbursts that eject mass require at least $E_{\rm min} \sim 5 \times 10^{46} \, {\rm erg}$ deposited in less than one dynamical time (such that a pressure pulse is formed). As expected, the corresponding mass loss is much smaller for compact stars (e.g., helium stars) due to their higher binding energy per unit mass, so a much larger energy budget is required for substantial mass ejection in those stars. Our results also indicate that the ejected mass is a sensitive function of the energy, scaling approximately as $m_{\rm ej} \propto E^{2.5}$ for $E > E_{\rm min}$, so low-energy outbursts can typically only unbind a very small fraction of the envelope unless $E \sim E_{\rm bind}$. 

It is notable that the energy scale of $\sim \! 10^{47} \, {\rm erg}$ required for shock-driven mass ejection is comparable to that predicted by the wave heating mechanism for pre-supernova outbursts \citep{quataert:12,shiode:14,fuller:17,fuller:18}. Our work only applies to wave heating in red supergiants, as wave heating in compact stars occurs slower than a dynamical time scale and results in continuum wind-driven mass loss rather than shock-driven mass loss. As discussed above, there is only a narrow energy range over which shocks drive partial mass loss from red supergiants, and this range is subtended by variations in wave energy heating due to different progenitor structures (Wu \& Fuller, in prep) and uncertainties stemming from the poorly constrained spectrum of convectively excited waves. Hence, within the range of present uncertainties, it is possible that many red supergiants suffer no eruptive pre-SN mass loss, while others suffer very large amounts of mass loss. 

We have focused on simple polytropic stellar models with constant opacity, but future work should examine how weak shocks affect more realistic stellar models. Our approximations are best for compact helium stars, in which the gas is nearly fully ionized, radiation pressure dominates (such that a constant $\gamma = 4/3$ is a good approximation), and for which electron scattering is the most important source of opacity near the photosphere (such that a constant $\kappa$ is a good approximation). Shock propagation through red supergiants may be significantly altered by the changing $\gamma$ due to partial hydrogen ionization, and the highly variable opacity, which varies by orders of magnitude in the outer layers due to the extreme sensitivity of H- opacity to temperature. Of particular importance is accounting for changes in $\gamma$ and $\kappa$ between the pre-shock and post-shock material, which are likely to be substantial for red supergiants, and which could significantly alter our quoted energy and mass estimates. 

Despite the limitations of our approach in the case of extended progenitors, we note that our results are in good agreement with those of \citealt{Kuriyama_Shigeyama_2020}. They have performed numerical radiation-hydrodynamics simulations of non-terminal explosions in a variety of progenitors, examining the resulting mass ejection and lightcurves. For their Wolf-Rayet and blue supergiant models, they find that the ejecta mass scales approximately as $m_{\rm ej} \propto E^{2.5}$, consistent with our numerical results. In these progenitors they have not they have not experimented with sufficiently small explosion energies to observe the existence of a minimal energy required for mass ejection. For their red supergiant models, they find that the amount of ejected mass rapidly decreases as the explosion energy decreases below roughly $10^{47} \rm \, erg$. We relate this sharp decline to the minimal explosion energy required for any mass ejection to occur, as the shock breaks out before accelerating to a fraction of the escape velocity.

The propagation of weak shocks in stellar envelopes has been studied analytically by \cite{coughlin_ro_quataert_2018}, who found a self-similar solution describing the propagation of a spherical weak shock wave in a hydrostatic medium with a point mass gravitational field, and later analyzed the solution's stability in \cite{coughlin_ro_quataert_2019} and \cite{ro_coughlin_quataert_2019}. Their solution exists when the density scales as $\rho \propto r^{-n}$, where $2<n<3.5$, with the shock propagating with a constant and order-unity Mach number, $v_{\rm sh} = V \sqrt{G M_0/r}$, where $V(n)\gtrsim 1$ is a constant and $M_0$ is the point mass dominating the gravitational field. In our polytropic models, the shock first weakens to $\mathcal{M} \approx 1$ near the center of the star where the density profile is roughly constant, too shallow for their self-similar solution to exist. The shock strength then increases near the stellar edge where the density rapidly decreases, scaling as a power law of the density from the surface. At this location, the density profile becomes too steep to sustain the \cite{coughlin_ro_quataert_2018} solution, so that self-similar behavior never appears in our polytropic models. Our models assumed a point explosion at the center of the star, but different energy injection mechanisms (such as the outward expansion during a failed supernova,  \citealt{coughlin_ro_quataert_2018}, or spatially extended wave heating in a stellar envelope) can allow for the self-similar solution to arise. As discussed in \cite{coughlin_ro_quataert_2019} and \cite{ro_coughlin_quataert_2019}, despite being weakly unstable, their self-similar solution is likely to prevail in the envelopes of supergiants where density scales as $\rho \propto r^{-n}$ over several orders of magnitude. As the weak shock approaches the stellar surface, it eventually transitions to a strong shock and starts accelerating, as dictated by the GFKS solution. By studying the transition between these two stages, the mass ejected from failed-supernovae explosions could be estimated, applying the concepts presented in this paper.


\section*{Acknowledgements}
IL thanks support from the Adams Fellowship. This research was partially supported by an ISF grant. JF acknowledges support from an Innovator Grant from The Rose Hills Foundation, and the Sloan Foundation through grant FG-2018-10515.

\section*{Data Availability}
The data underlying this article will be shared on reasonable request to the corresponding author.



\bibliographystyle{mnras}
\bibliography{shockmassloss,CoreRotBib}



\appendix

\section{Weak planar shock in power-law density profile} \label{sec:app_weak_planar_shock}
Consider a sound pulse propagating in a medium whose density varies as a power-law of the distance measured from the edge. After steepening to a weak shock, the pulse width and amplitude evolve due to two different effects discussed in section \ref{sec:weak_planar_shock} - the pulse's finite width, and non-linear steepening and widening.

We quantify the relative importance of the two effects with the dimensionless number $Z=(\delta v \, x)/(c_s \Delta x)$, where $\delta v$ is the material velocity amplitude, $x$ is the distance from the edge, $c_s$ is the local sound speed at $x$, and $\Delta x$ is the pulse's width.

Here we show that $Z$ always evolves towards a value of order unity, where the two dispersive effects are in balance. To that end, we express the energy per unit surface area carried by the pulse
\begin{equation}
    E \approx \rho \, \Delta x \, \delta v^2 \,,
\end{equation}
and rewrite
\begin{equation}
    Z \approx \left( \frac{E x^2}{\rho \, c_s^2 \, \Delta x^3} \right)^{1/2} \,.
\end{equation}

Taking the full derivative with respect to $x$, we get
\begin{equation} \label{eq:dZdX_gen}
    \frac{1}{Z} \frac{dZ}{dx} = \frac{1}{2E} \frac{\partial E}{\partial x} + \left( \frac{1-n}{2} \right) \frac{1}{x} - \frac{3}{2 \Delta x} \frac{d(\Delta x)}{dx} \,,
\end{equation}
where we used the fact $\rho \propto x^n$ and $c_s \propto x^{1/2}$.

We first consider the limit $Z \ll 1$. Here the pulse width decreases as it approaches the surface as
\begin{equation}
    \frac{d(\Delta x)}{dx} = \frac{\Delta x}{2 x} \,.
\end{equation}
The pulse energy can only decrease as it approaches the surface, and we assume $E \propto x^{\varepsilon_E}$, where $\varepsilon_E > 0$. Plugging in equation \ref{eq:dZdX_gen} we get
\begin{equation}
    \frac{d \log{Z}}{d \log{x}} = -\left( \frac{n}{2} + \frac{1}{4} - \frac{\varepsilon_E}{2} \right) \,,
\end{equation}
thus, as long as $\varepsilon_E < (n+1/2)$, we find that if initially $Z \ll 1$, its value increases as a power law as the pulse approaches the surface (decreasing $x$).

In the complementary limit, $Z \gg 1$, $\Delta x$ increases as the pulse approaches the surface as
\begin{equation}
    \frac{d (\Delta x)}{dx} = -\frac{\gamma+1}{2} \frac{\delta v}{c_s} \,.
\end{equation}
Plugging into equation \ref{eq:dZdX_gen} and neglecting sub-leading terms
\begin{equation}
    \frac{d \log{Z}}{d \log{x}} = \frac{3(\gamma+1)}{4} Z \,,
\end{equation}
where we used $Z \gg 1$. Therefore, in this limit, $Z$ decreases as the pulse approaches the edge of the medium. Put together,
\begin{equation}
    \frac{d \log{Z}}{d \log{x}} =
    \begin{cases}
      < 0 & Z \ll 1\\
      > 0 & Z \gg 1\\
    \end{cases} \,,
\end{equation}
and thus the pulse evolves towards a state with $Z \sim 1$.


\bsp	
\label{lastpage}
\end{document}